\documentstyle[12pt]{article}
\begin{document}
\begin{center}
{\bf Relations among QCD corrections beyond leading order
\footnote{Work supported in part by EU contract FMRX-CT98-0194}} \\
\end{center}
\begin{center}{\bf J. Bl\"umlein, V. Ravindran and W.L. van Neerven 
\footnote {On leave of absence from Instituut-Lorentz, 
University of Leiden, P.O. Box 9506, 2300 HA Leiden, The Netherlands.} }
\end{center}
\begin{center}
DESY-Zeuthen, Platanenallee 6, D-15738 Zeuthen, Germany.
\end{center}
  
\noindent Symmetries are known to be a very useful guiding tool to understand the 
dynamics of various
physical phenomena.  Particularly, continuous symmetries played
an important role in particle physics to unravel the structure of dynamics
at low as well as high energies.  In hadronic physics, such symmetries
at low energies were found to be useful to classify various hadrons.  
At high energy, where the masses of the particles can be neglected, one
finds in addition to the above mentioned symmetries
new symmetries such as conformal and scale invariance. This for instance
happens in deep inelastic lepton-hadron scattering (DIS) where
the energy scale is much larger than the hadronic mass scale.
At these energies one can in principle ignore the mass scale and the resulting
dynamics is purely scale independent \cite{callan}. 

\noindent We first discuss the Drell-Levy-Yan relation (DLY) \cite{drell} which relates 
the  structure  functions 
$F(x,Q^2)$ measured in deep inelastic scattering to the fragmentation functions 
$\tilde F(\tilde x,Q^2)$ observed 
in $e^+~e^-$-annihilation. Here $x$ denotes the Bj{\o}rken
scaling variable which in deep inelastic scattering and $e^+~e^-$-annihilation
is defined by $x=Q^2/2p.q$ and $\tilde x=2p.q/Q^2$ respectively.
Notice that in deep inelastic scattering the virtual photon
momentum $q$ is spacelike i.e. $q^2=-Q^2<0$ whereas in $e^+~e^-$-annihilation it
becomes timelike $q^2=Q^2>0$. Further $p$ denotes the in or outgoing hadron
momentum.
The DLY relation looks as follows 
\begin{eqnarray}  
\label{eq24}
\tilde F_i(\tilde x,Q^2)= x {\cal A }c\left[F_i(1/x,Q^2)\right] \,,
\end{eqnarray}
where ${\cal A }c$ denotes the analytic continuation from the region
$0<x \le 1$ (DIS) to $1<x<\infty$ (annihilation region). 
At the level of splitting functions we have
\begin{eqnarray}  
\label{eq25}
\tilde P_{ij}(\tilde x)= x {\cal A }c\left[P_{ji}(1/x)\right] \,.
\end{eqnarray}
At LO, one finds  $\tilde P^{(0)}_{ij}(\tilde x)=P^{(0)}_{ji}(x)$, for $ x < 1$ 
which is nothing but Gribov-Lipatov relation \cite {gribov}.  This relation
in terms of physical observables is known to be violated when one goes beyond leading order.  
There is a similar violation
of the DLY relation among coefficient functions as well.
The DLY (analytical continuation) relation defined above holds at the level of
physical quatities provided the analytical continuation is performed in both $x$ as well as
the scale $Q^2$ ($Q^2 \rightarrow -Q^2$).  For example
the $\Gamma_{ij}$ appearing in the following physical observables \cite{bluemlein}
\begin{displaymath}
Q^2{\partial \over \partial Q^2}\left( \begin{array}{c}
F_A\\
{F_B}
\end{array} \right)= 
\left( \begin{array} {cc}
\Gamma_{AA} & \Gamma_{AB}\\
\Gamma_{BA} & \Gamma_{BB}
\end{array} \right)
\left( \begin{array}{c}
F_A\\
{F_B}
\end{array} \right)
\end{displaymath}
satisfy the DLY relation,  where $F_A,F_B$ are any two structure functions
and $Q$ is the scale involved in the process. 
The violation of the DLY relation for the splitting functions
and the coefficient functions is just
an artifact of the adopted regularization method and the subtraction scheme.
When these coefficient functions are combined with the splitting functions
in a scheme invariant way, as for instance happens for the structure
functions  and fragmentation functions the DLY relation holds.
The reason for the cancellation of the DLY violating terms among 
the splitting functions and coefficient functions is that the former
are generated by simple scheme transformations.  

We now discuss Supersymmetric relations among splitting functions 
which determine the evolution of quark and gluon parton densities. These
relations are valid when QCD becomes a supersymmetric 
${\cal N}=1$ gauge field theory where both quarks and gluons are put
in the adjoint representation with respect to the local gauge symmetry
$SU(N)$. In this case one gets a simple relation between the colour factors
which become $C_F=C_A=2 T_f = N$.
In the case of spacelike splitting functions, which determine
the evolution of the parton densities in deep inelastic lepton-hadron
scattering, one has made the claim (see \cite{dokshitser}) that the
combination defined by 
\begin{eqnarray}
\label{eq15}
{\cal R}^{(i)}= P_{qq}^{(i)}- P_{gg}^{(i)}+ P_{gq}^{(i)}- P_{qg}^{(i)} \,,
\end{eqnarray}
is equal to zero, i.e., ${\cal R}^{(i)}=0$. This relation should follow
from an ${\cal N}=1$ supersymmetry although no proof has been given
yet. An explicit calculation at leading order(LO) confirms this claim so
that we have ${\cal R}^{(0)}=0$.
However at next to leading order(NLO), when these splitting functions are 
computed in the ${\overline {\rm MS}}$-scheme, it turns out that 
${\cal R}_{\overline {\rm MS}}^{(1)}\not =0$. 
The reason that this relation is violated can be attributed to the 
regularization method and the 
renormalization scheme in which these splitting functions are computed.
In this case it is $D$-dimensional regularization and the $\overline{\rm MS}$-scheme 
which breaks the supersymmetry.  In fact, the breaking occurs already at the
$\epsilon$ dependent part of the leading order splitting functions. Although
this does not affect the leading order splitting functions in the limit 
$\epsilon \rightarrow 0$ it leads to
a finite contribution at the NLO level via the $1/\epsilon^2$ terms which
are characteristic of a two-loop calculation.
If one carefully removes such breaking terms at the LO level consistently,
one can avoid these terms at NLO level. They can also be
avoided if one uses $D$-dimensional reduction which preserves the supersymmetry. 
The above observations also apply to the timelike splitting functions, 
which determine the evolution of fragmentation functions.

\end{document}